\documentclass[reprint,twocolumn,aps,prb,superscriptaddress]{revtex4-2}
\usepackage[T1]{fontenc}
\usepackage[utf8]{inputenc}
\usepackage{lmodern}
\usepackage{graphicx}
\usepackage{amsmath}
\usepackage{xcolor}
\usepackage{float}
\usepackage{hyperref}
\usepackage{braket}
\usepackage{amssymb,amsmath,amsthm, array}
\usepackage{mdframed}
\usepackage{systeme,mathtools}
\usepackage{yfonts}
\usepackage{comment}

\DeclareUnicodeCharacter{0301}{-}

\begin{document}
\renewcommand{\vec}[1]{\mathbf{#1}}
\newcommand{\ii}{\mathrm{i}}
\def\ya#1{{\color{orange}{#1}}}

\title{Relativistic quantum entanglement in a bipartite charged scalar system}

\author{T. Nadareishvili}

\affiliation{Faculty of Exact and Natural Sciences Chavchavadze av.3, 0128 Tbilisi, Georgia}
\affiliation{Inst. of High Energy Physics, Iv. Javakhishvili Tbilisi State University, University Str. 9, 0109, Tbilisi, Georgia}

\author{S. Stagraczy\'nski}
\affiliation{Department of Physics and Medical Engineering, Rzesz\'ow University of Technology, 35-959 Rzesz\'ow, Poland}

\author{L. Chotorlishvili}
\affiliation{Department of Physics and Medical Engineering, Rzesz\'ow University of Technology, 35-959 Rzesz\'ow, Poland}

\begin{abstract}
Continuous  variable entanglement in a system of two interacting charged scalar mesons is studied.
The mesons mutual interaction   is mediated by a central symmetric Coulomb potential.
We work out the  difference between relativistic and non-relativistic limits on the basis of
 the  continuous  variable separability criteria  and prove rigorously the following theorem for our system:
While a bipartite charged scalar mesonic quantum system   with centrally symmetric interaction  is entangled in the
relativistic  regime,  its non-relativistic counterpart can be separable with non entanglement.

\end{abstract}
\maketitle

\textit{Introduction}:-
Quantum entanglement is a phenomenon inherent  to quantum correlations and
with no classical analogy \cite{Horodecki,Wootters,Zurek,Osterloh,Braunstein,Harshit,Rohit,Abhishek,Podist}.
As approaching the  classical  limit the entanglement in a quantum system  disappears gradually.
This applies to systems depending on either  discrete or/and continuous  variables
\cite{Simon,Weedbrook,Silberhorn,Adesso,Toscano,Serafini,Shchukin,Shchukin,Werner}.
The concept of entanglement  allows an insight into the fundamentals of quantum physics
\cite{Horodecki,Wootters,Zurek,Osterloh,Braunstein,Simon,Weedbrook,Silberhorn,Adesso,Toscano,Serafini,Shchukin,Shchukin,Werner}
with prospect applications  for long-distance quantum computation in  optics and  optical microcavities
\cite{Knill,Duan,Vahala}, superconducting circuits \cite{Nori1,Nori2},
quantum computing using molecular magnets
\cite{Leuenberger}, nitrogen vacancy centers \cite{Mishra} and also in quantum thermodynamics \cite{Schmidt-Kaler,Abah,Lutz,Deffner,Talkner}.
On the fundamental side,  quantum system may be relativistic or nonrelativistic.
While the essence of entanglement of non-relativistic quantum systems is widely explored, some aspects of purely
 quantum correlations in the relativistic regime are still to be fully uncovered.
 The dynamics of  charged scalar particles such as  the $\pi^{\pm},~\pi^{0}$ mesons,   is generally well captured by  the Klein-Gordon (KG)) equation.
  For  distances larger than the Yukawa radius $r_{0}=10^{-15}$m and energies smaller than 140 Mev the nuclear forces are ignorablly  small in which case a bipartite system of such two mesons can be described by the KG equation including a central symmetric Coulomb potential \cite{Bogoliubov}.
We note that mesons of the $J/\psi$ and $\gamma$ families can be treated through the nonrelativistic Schr\"odinger equation, as well.   The (nonrelativistic) Schr\"odinger equation does not yield however the correct spectrum of ordinary mesons forming the light quarks (cf. Ref.\onlinecite{Lichtenberg} for more details). \\
The present work is devoted to  the study of the continuous  variable entanglement in a bipartite system of the two interacting mesons $\big(\pi^{+}, \pi^{-}\big)$.
Using the KG equation with central symmetric interaction we were able to obtain analytical results.
The question of  primary interest is the role of relativity  in quantum correlations in a system with
a long-range central symmetric potential.
 We proof that systems that are disentangled in non-relativistic limit can be entangled in a relativistic regime.\\
\textit{ Continuous   variable entanglement  in KG systems}:-
To quantify continuous   variable entanglement we utilize the method developed in Ref. \onlinecite{Zoller}, showing  that
a state of a bipartite quantum system is separable if  inequality holds
\begin{equation}
\begin{aligned}
\big<\big(\Delta \hat{u}\big)^{2}\big>+\big<\big(\Delta \hat{v}\big)^{2}\big> \geq a^{2}+1/a^{2}.  \\
\end{aligned}
\label{criteria}
\end{equation}
 $\Delta \hat{u}$ and $\Delta \hat{v}$ are the variances of  respectively the operators $\hat{u}=ar_{1}+1/ar_{2}$ and $\hat{v}=a\hat{p}_{1}-1/a\hat{p}_{2}$,
$a$ is an arbitrary finite real number and $r_{1,2}$, $\hat{p}_{1,2}$ are the operators of the canonical coordinate and momentum.

We note that Eq.(\ref{criteria}) is the sufficient criteria. When Eq.(\ref{criteria}) holds for the arbitrary nonzero finite real number $a$, the state is separable.
However  some separable states may violate condition Eq.(\ref{criteria}). Therefore the sufficient criteria should be supplemented by necessary criteria:
For inseparable states, the total variance is required \cite{Zoller} to be larger or equal to:
\begin{equation}
\begin{aligned}
\big<\big(\Delta \hat{u}\big)^{2}\big>+\big<\big(\Delta \hat{v}\big)^{2}\big> \geq |a^{2}-1/a^{2}|.  \\
\end{aligned}
\label{sufficientcriteria}
\end{equation}
Thus if Eq.(\ref{criteria}) is violated and Eq.(\ref{sufficientcriteria}) holds one can precisely argue that state is entangled and vice versa.
For self-consistency, we explore both criteria. We adopt standard notations for four-vectors:
$\hat{p}_{\mu}=\big(E/c,-\vec{\hat{p}}\big),~~~x_{\mu}=\big(ct,-\vec{x}\big)$ and $\sum_{\mu=0}^{3}\hat{p}_{\mu}\hat{p}^{\mu}=E^{2}/c^{2}-\vec{\hat{p}}^{2}=m^{2}c^{2}$.
Then the (relativistic) Klein-Gordon equation of two mesons in a central symmetric interaction potential read \cite{Lichtenberg}
\begin{equation}
\begin{aligned}
\big\{\hat{p}^{2}+m^{2}\big\}\Psi \big(r_{1},r_{2}\big)=\frac{1}{4}\big(M-V\big(r_{1}-r_{2}\big)\big)^{2}\Psi \big(r_{1},r_{2}\big).
\end{aligned}
\label{equation}
\end{equation}
Here $M-V\big(r_{1}-r_{2}\big)=\big(\hat{p}^{2}+m^{2}_{1}\big)^{1/2}+\big(\hat{p}^{2}+m^{2}_{1}\big)^{1/2}$ is the total energy of the system (velocity of light $c=1$ and $\hbar=1$).
In the relativistic center-of-mass system \cite{Kang}: $\hat{p}_{1}=-\hat{p}_{2}=\hat{p}$, $r_{1}=E_{2}/(E_{1}+E_{2})r$,
$r_{2}=-E_{1}/(E_{1}+E_{2})r$, $r=r_{1}-r_{2}$, the energies of the free particles $E_{1}=\big(\hat{p}^{2}+m^{2}_{1}\big)^{1/2}$,
$E_{2}=\big(\hat{p}^{2}+m^{2}_{1}\big)^{1/2}$. The mass of the free particles we set to be equal $m_{1}=m_{2}=m$.
We divide  the KG equation by $m_{\mu}c^{2}$
where $m_{\mu}$ is the meson mass and use  a  dimensionless unit of distance $\vec{r}\rightarrow\vec{r}/r_{0}$, where $r_{0}$ is the Yukawa radius  (note also  $m=1$). \\
Integrals of the form $\big<\Psi|r_{1,2}|\Psi\big>$ and $\big<\Psi|\hat{p}_{1,2}|\Psi\big>$ in Eq.(\ref{criteria}) vanish. Therefore,
for calculations of the variances of the operators $\Delta \hat{u}$ and $\Delta \hat{v}$ we need only the radial part of the wave function
$\Psi \big(r_{1},r_{2}\big)$ see Eq.(\ref{equation}). After somewhat involved calculations we deduce from  Eq.(\ref{criteria})

\begin{equation}
\begin{aligned}
\bigg(a^{2}+\frac{1}{a^{2}}\bigg)B-\frac{\big<\big(r_{1}-r_{2}\big)^{2}\big>}{2}+D>a^{2}+\frac{1}{a^{2}}.
\end{aligned}
\label{shortcriteria}
\end{equation}
Here  we introduced the notations
\begin{eqnarray}
&& B={\frac{\big<\big(r_{1}-r_{2}\big)^{2}\big>}{4}}+4\bigg[\bigg(\frac{M^{2}}{4}-m^{2}\bigg)+ \nonumber\\
&& \quad \frac{1}{4}\big<V^{2}\big(r_{1}-r_{2}\big)\big>-\frac{M}{2}\big<V\big(r_{1}-r_{2}\big)\big>\bigg],
\label{Bshort}
\end{eqnarray}
and
\begin{eqnarray}
&& D=-8\bigg\{\bigg(\frac{M^{2}}{4}-m^{2}\bigg)+ \frac{1}{4}\big<V^{2}\big(r_{1}-r_{2}\big)\big>-\nonumber\\
&& \quad \frac{M}{2}\big<V\big(r_{1}-r_{2}\big)\big>\bigg\}.
\label{Dshort}
\end{eqnarray}
To further analyze the separability criteria, the exact expectation values of the operators are required  that enter  Eq.(\ref{shortcriteria}).
Before embarking on this task we perform  a qualitative asymptotic analyses for the cases $a=1$ and $a\gg1$, $a\ll1$. After some algebra we infer from
Eq.(\ref{shortcriteria}) for the asymptotic cases $a\ll1$ and $a\gg1$
\begin{equation}
\begin{aligned}
A\big(r_{1}-r_{2}\big)+\frac{1}{4}\big<\big(r_{1}-r_{2}\big)^{2}\big>-1>0,
\end{aligned}
\label{anonequalone}
\end{equation}
and for  $a=1$
\begin{equation}
\begin{aligned}
0\cdot A\big(r_{1}-r_{2}\big)+0\cdot\frac{\big<\big(r_{1}-r_{2}\big)^{2}\big>}{4}-2>0.
\end{aligned}
\label{aequalone}
\end{equation}
where
\begin{eqnarray}
&& A\big(r_{1}-r_{2}\big)=4\bigg\{\bigg(\frac{M^{2}}{4}-m^{2}\bigg)+\frac{1}{4}\big<V^{2}\big(r_{1}-r_{2}\big)\big>-\nonumber\\
&& \quad \frac{M}{2}\big<V\big(r_{1}-r_{2}\big)\big>\bigg\}.
\label{Bnotation}
\end{eqnarray}
From Eq.(\ref{aequalone}) follows  for $a=1$ the separability condition never holds. For $a\gg1,~~a\ll1$ there may exist a finite  expectation value
of $\big<r^{2}\big>=\big<\big(r_{1}-r_{2}\big)^{2}\big>$ such that $\big<r^{2}\big>\gg 4\big(1-A\big(\big<r\big>\big)\big)$. Obviously, the expectation value of $\big<r^{2}\big>$ increases with the radial quantum number and for  some  excited states may reach the critical value
$\big<r^{2}_{c}\big>=4\big(1-A\big(\big<r_{c}\big>\big)\big)$. However, even in this excited state Eq.(\ref{aequalone}) fails to hold.
These  observations are general allowing for the following statement:
For a bipartite quantum system with a central symmetric interaction potential, the separability condition   does not holds  for $a=1$,
while for $a\gg1$ and $a\ll1$ the separability may occurs depending on whether the system is relativistic or not.
Fact that inequality does not hold for particular value of the parameter $a=1$ is not enough to constitute the existence of the entangled state.
Note that according to the necessary and sufficient theorem entangled states violate inequality for all real values of $a$ and not only for $a=1$.\\

Proceeding further to  explore the  properties of  Eq.(\ref{equation}) with the symmetric coulomb potential $V=-\alpha/|r_{1}-r_{2}|$
(here $\alpha\rightarrow \alpha r_{0}/m_{\mu}c^{2}$ is a dimensionless interaction constant) we have to calculate the following type of integrals
for  obtaining  explicit expressions for the parameters $B$ and $D$  (cf.  Eqs.(\ref{Bshort},\ref{Dshort})

$$I_{-1}=\alpha\big<\Psi|\big(r_{1}-r_{2}\big)^{-1}|\Psi\big>,$$
$$I_{-2}=\alpha^{2}\big<\Psi|\big(r_{1}-r_{2}\big)^{-2}|\Psi\big>,$$ and
$$I_{2}=\big<\Psi|\big(r_{1}-r_{2}\big)^{2}|\Psi\big>.$$
The solution of Eq.(\ref{equation}) can be found in the following form
\begin{eqnarray}
&& R=C\rho^{-1/2+p}e^{-\rho/2}F\bigg(1/2+\xi-\lambda, 1+2\xi, \rho\bigg).
\label{intermediate1}
\end{eqnarray}
Here $F\big(1/2+\xi-\lambda, 1+2\xi, \rho\big)$ is the Hypergeometric function.
Other parameters are defined as: $k=\sqrt{4m^{2}-M^{2}}$, $\lambda=M\alpha/2k$, $\rho=kr$, $\xi=\sqrt{\big(l+1/2\big)^{2}-\alpha^{2}/4}$.\\
To find the normalization constant $C$ from the equation $\int^{\infty}_{0}R^{2}r^{2}dr=1$ we utilize the following form of the hypergeometric function
\begin{eqnarray}
&& F\big(-n,s+1,x\big)=Q^{s}_{n}\big(x\big)\frac{\Gamma\big(s+1\big)}{\Gamma\big(s+n+1\big)},
\label{intermediate2}
\end{eqnarray}
where $Q^{s}_{n}\big(x\big)$ is the generalized Laguerre polynomial and $\Gamma\big(s+1\big)$,  $\Gamma\big(s+n+1\big)$ are the Gamma functions.
Taking into account that
\begin{eqnarray}
&& \int^{\infty}_{0}e^{-x}x^{\alpha+k}\big[Q_{n}^{\alpha}\big(x\big)\big]dx=\frac{\Gamma\big(\alpha+k+1\big)\Gamma\big(\alpha+n+1\big)}{n!\Gamma\big(\alpha+1\big)}
\nonumber\\
&& \quad {}_{3}F_{2}
\left(
\begin{array}{c}
-k,k+1,-n\\
1,\alpha + 1
\end{array}
\right),
\label{intermediate3}
\end{eqnarray}
for the solution of Eq.(\ref{equation}) we deduce
\begin{eqnarray}
&& \big|\Psi \big(r_{1},r_{2}\big)\big>_{n}=\frac{k^{3/2}}{\sqrt{n!\big(2n+2\xi+1\big)}}\frac{\sqrt{\Gamma\big(2\xi+1+n\big)}}{\Gamma\big(2\xi+1\big)}\times \nonumber\\
&& \quad \rho^{-1/2+\xi}\exp\big(-\rho/2\big)F\big(1/2+\xi-\lambda, 1+2\xi, \rho\big).
\label{solutionwavefunction}
\end{eqnarray}
Skipping technical details the final results read
\begin{equation}
\begin{aligned}
I_{-2}=\frac{\alpha^{2} C^{2}\Gamma^{2}\big(2\xi+1\big)}{\Gamma\big(2\xi+1+n\big)}\frac{n!}{2\xi}\frac{1}{k},
\end{aligned}
\label{I-2}
\end{equation}
\begin{equation}
\begin{aligned}
I_{-1}=\frac{\alpha C^{2}}{k^{2}}\frac{\Gamma^{2}\big(2\xi+1\big)}{\Gamma\big(2\xi+1+n\big)}n!,
\end{aligned}
\label{I-1}
\end{equation}
and
\begin{equation}
\begin{aligned}
I_{2}=\frac{C^{2}}{k^{5}}\frac{\Gamma^{2}\big(2\xi+1\big)}{\Gamma\big(2\xi+1+n\big)}n!F_{0}.
\end{aligned}
\label{I2}
\end{equation}
Here
\begin{eqnarray}
&& F_{0}=\big(2\xi+1\big)\big(2\xi+2\big)\big(2\xi+3\big)\bigg\{1+\frac{2n}{\big(2\xi+1\big)}+\nonumber\\
&& \quad \frac{10n\big(1-n\big)}{\big(2\xi+1\big)\big(2\xi+2\big)}+\nonumber\\
&& \quad \frac{20n\big(1-n\big)\big(2-n\big)}{\big(2\xi+1\big)\big(2\xi+2\big)\big(2\xi+3\big)}\bigg\},
\label{I2}
\end{eqnarray}
\begin{equation}
\begin{aligned}
C^{2}=\frac{k^{3}}{n!\big(2n+2\xi+1\big)}\frac{\Gamma\big(2\xi+1+n\big)}{\Gamma^{2}\big(2\xi+1\big)},
\end{aligned}
\label{cznachenie}
\end{equation}
 and $n,l$ are the radial and the orbital quantum
numbers. \\
Accounting for  Eqs.(\ref{I-2}-\ref{cznachenie}), for Eq.(\ref{Bshort}) and Eq.(\ref{Dshort}) we deduce
\begin{eqnarray}
&& B=\frac{C^{2}\Gamma^{2}\big(2\xi+1\big)n!}{\Gamma\big(2\xi+n+1\big)}\bigg\{\frac{F_{0}}{4k^{5}}+\frac{\alpha^{2}}{2\xi k}+\frac{2M\alpha}{k^{2}}\bigg\}+ \nonumber\\
&& \quad \big(M^{2}-4m^{2}\big),
\label{Bsimple}
\end{eqnarray}
and
\begin{eqnarray}
&& D=-\frac{4\alpha C^{2}\Gamma^{2}\big(2\xi+1\big)n!}{\Gamma\big(2\xi+n+1\big)}\bigg\{\frac{\alpha}{4\xi k}+\frac{M}{k^{2}}\bigg\}- \nonumber\\
&& \quad 8\big(M^{2}/4-m^{2}\big).
\label{Dsimple}
\end{eqnarray}
The obtained expressions Eq.(\ref{Bsimple}), Eq.(\ref{Dsimple}) for $B$ and $D$ are general and valid also in the relativistic regime.
 We utilize them when analyzing
the separability criteria Eq.(\ref{shortcriteria}).
The relativistic character of the problem impacts the mass of the system. In particular, the
mass of the system of two interacting meson has the form
\begin{eqnarray}
&& M=\frac{2m}{\sqrt{1+\alpha^{2}/4N^{2}}}.
\label{Massseries}
\end{eqnarray}
Here  $N=n+1/2+\xi$ and $n=0,1,2...$ is the main quantum number.
For simplicity, we  analyze the separability
criteria Eq.(\ref{shortcriteria}) for two particular cases: $a\gg1,~a\ll1$.
Taking into account Eq.(\ref{Bsimple}), Eq.(\ref{Dsimple}), Eq.(\ref{Massseries}) for the separability
criteria in the relativistic limit we find
\begin{eqnarray}
&& \frac{\alpha}{N}\frac{m^{2}\alpha^{2}}{N^{2}+\alpha^{2}/4}\bigg\{\frac{\alpha}{4\xi}+\frac{2N}{\alpha}\bigg\}-\frac{m^{2}\alpha^{2}}{N^{2}+\alpha^{2}/4}+  \nonumber\\
&& \quad \frac{1}{8N}\frac{N^{2}+\alpha^{2}/4}{m^{2}\alpha^{2}}F_{0}-1>0.
\label{criteriarelativistic}
\end{eqnarray}
Non-relativistic separability criteria  can also  be derived straightforwardly as
\begin{eqnarray}
&& \frac{m^{2}\alpha^{2}}{N_{1}^{2}}+\frac{1}{2}\frac{N_{1}^{2}}{m^{2}\alpha^{2}}\big[5N_{1}^{2}+1-3l\big(l+1\big)\big]-1>0,  \nonumber\\
&& N_{1}=n+l+1.
\label{noncriteriarelativistic}
\end{eqnarray}
The analysis  of Eq.(\ref{noncriteriarelativistic}) shows that for $a\gg 1,~a\ll1$ the  non-relativistic inequality holds always,
while for $a=1$ it never holds see Eq.(\ref{aequalone}). On the other hand, for $l=0,~~\alpha=1/2,~~n=2,3$ and $l=1,~n=3$ for $\alpha \simeq 2.25$
the relativistic inequality Eq.(\ref{criteriarelativistic}) is violated and  the states are entangled.
Thus,  in contrast  to the non-relativistic case,  the separability criteria in the relativistic case are rigorously violated and state is entangled.\\
\textit{General separability criteria for KG systems}:-
With this  we proceed now to explore the separability criteria in the general case for arbitrary $a$
\begin{eqnarray}
&&
Y_{LHS}>Y_{RHS},\nonumber\\
&& Y_{LHS}=\big(a^{2}+1/a^{2}\big)\times\nonumber\\
&&\bigg\{\frac{1}{2n+2\xi+1}\frac{4\alpha^{2}}{4N^{2}+\alpha^{2}}\bigg[\frac{F_{0}}{4}\frac{\big(4N^{2}+\alpha^{2}\big)^{2}}{16\alpha^{4}}+  \nonumber\\
&& \frac{\alpha^{2}}{2\xi}+4N\bigg]-\frac{4\alpha^{2}}{4N^{2}+\alpha^{2}}\bigg\}-\frac{4N^{2}+\alpha^{2}}{8\alpha^{2}}\frac{F_{0}}{\big(2n+2\xi+1\big)}- \nonumber\\
&& \frac{4\alpha}{2n+2\xi+1}\frac{4\alpha^{2}}{4N^{2}+\alpha^{2}}\bigg\{\frac{\alpha}{4\xi}+\frac{2N}{\alpha}\bigg\}+\frac{32\alpha^{2}}{4N^{2}+\alpha^{2}}, \nonumber\\
&&Y_{RHS} =\big(a^{2}+1/a^{2}\big).\nonumber\\
\label{separabilitynumeric}
\end{eqnarray}
The separability criteria in the general case Eq.(\ref{separabilitynumeric}) are rather involved and
 can be fully accessed  only numerically. Numerical analysis shows that for  different values of the interaction
constant $\alpha$ and the parameter $a$, the inequality Eq.(\ref{separabilitynumeric}) is violated for different quantum states.
However, the entanglement the criteria impose a violation of the inequality
Eq.(\ref{separabilitynumeric}) for arbitrary positive $a>0$. The only states that meet such a requirement are the relativistic states $l=0,~n=2,3$ and $l=1,~n=3$ for $\alpha \simeq 2.25$.

\begin{figure}[ht]
	\includegraphics[width=0.9\columnwidth]{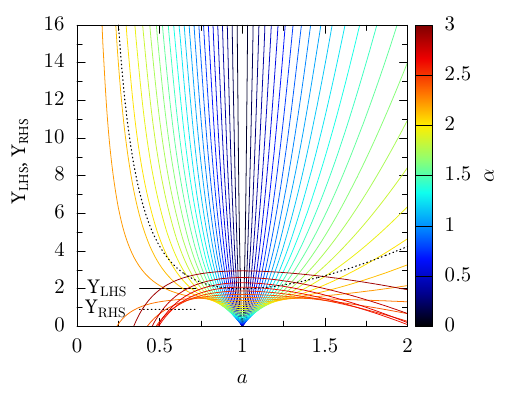}
	\caption{The functions $Y_{RHS}$  (single dotted line) and $Y_{LHS}$  (set of solid lines), as expressed in Eq.(\ref{separabilitynumeric}) are plotted for different values of $\alpha$ and $a$ for the quantum state $l = 1$, $n=3$.
For a given quantum state, $Y_{RHS}$  depends only on the parameter $a$, while $Y_{LHS}$ depends on $\alpha$, as well. The system is entangled if $Y_{RHS}$  exceeds $Y_{LHS}$  which is evidently  the case for
several values of $\alpha$.}
\label{Fig.1}
\end{figure}

\begin{figure}[ht]
	\includegraphics[width=0.9\columnwidth]{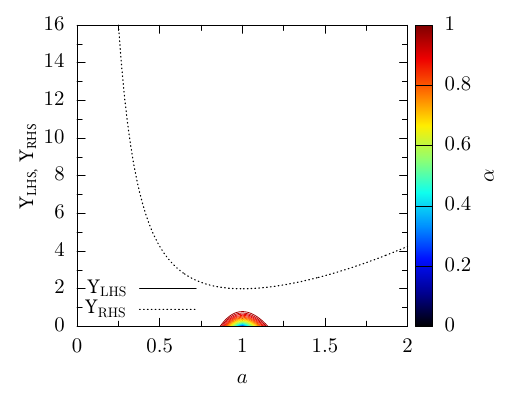}
	\caption{ Same notation and functions as in \ref{Fig.1} but here we consider the case
		the quantum state $l = 0$, $n=2$.
Figures shows that  the system is entangled  ( $Y_{RHS}>Y_{LHS}$)  for $0<\alpha$<1.}
\label{Fig.2}
\end{figure}

Results of numerical analyze of Eq.(\ref{separabilitynumeric}) are presented in Fig.(\ref{Fig.1}) and Fig.(\ref{Fig.2}).

\textit{Conclusions:}- The so-far obtained results allow for the following theorem:
 While a bipartite KG relativistic quantum  system with a central symmetric interaction potential is entangled,
the non-relativistic counterpart can be disentangled.

\end{document}